\documentclass[reprint,amsmath,amssymb,aps]{revtex4-2}
\usepackage{svg}
\usepackage{graphicx}
\usepackage{dcolumn}
\usepackage{bm}
\usepackage{hyperref}
\usepackage{csquotes}
\begin{document}
\title{Enhanced diffusion over a periodic trap by hydrodynamic coupling to an elastic mode}
\author{Juliette Lacherez$^1$}
\author{Maxime Lavaud$^{1,2}$}
\author{Yacine Amarouchene$^1$}
\email{yacine.amarouchene@u-bordeaux.fr}
\author{David S. Dean$^1$}
\email{david.dean@u-bordeaux.fr}
\author{Thomas Salez$^1$}
\email{thomas.salez@cnrs.fr}
\affiliation{$^1$Univ. Bordeaux, CNRS, LOMA, UMR 5798, F-33400, Talence, France}
\affiliation{$^2$Univ. Bordeaux, CNRS, Bordeaux INP, CBMN, UMR 5248, F-33600, Pessac, France}
\date{\today}
\begin{abstract}
In many physical systems, degrees of freedom are coupled \emph{via} hydrodynamic forces, even in the absence of Hamiltonian interactions. A particularly important and widespread example concerns the transport of microscopic particles in fluids near deformable boundaries. 
In such a situation, the influence of elastohydrodynamic couplings on Brownian motion remains to be understood. Unfortunately, the temporal and spatial scales associated with the 
thermal fluctuations of usual surfaces are often so small that their deformations are difficult to monitor experimentally, together with the much slower and larger particle motion at stake. Here, we propose a minimal model describing the hydrodynamic coupling of a colloidal particle to a fluctuating elastic mode, in presence of an external periodic potential. We demonstrate that the late-time diffusion coefficient of the particle increases with the compliance of the elastic mode. Remarkably,
  our results reveal that, and quantify how: i) spontaneous microscopic transport in complex environnements can be affected by soft boundaries -- a situation with numerous practical implications in nanoscale and biological physics; ii) the effects of fast and tiny surface deformations are imprinted over the long-term and large-distance colloidal mobility -- and are hence measurable in practice.
 \end{abstract}

\maketitle
Brownian motion refers to the random movement of a colloid at equilibrium, induced by its collisions with the molecules of the surrounding fluid. It
 has played a foundational role in modern statistical physics since its 
 discovery~\cite{brown_xxvii_1828} and theoretical description~\cite{einstein_uber_1905,von_smoluchowski_zur_1906} over a century ago. 
 The celebrated experiments of Perrin on colloids in a gravitational field~\cite{perrin_mouvement_1909} validated further the theoretical description and confirmed the atomic nature of matter. Extending the classical framework of Brownian motion in bulk Newtonian fluids to more realistic systems, such as particles diffusing in complex environnements, opens up new challenges for nanoscale and biological physics. One common route is to study Brownian motion in viscoelastic fluids or intracellular media, as typically performed through microrheology~\cite{Mackintosh1999}, which revealed the emergence of anomalous diffusion~\cite{Metzler2000}. Another source of complexity lies in the introduction of topological constraints and boundaries. For instance, a simple-yet-canonical situation where interfaces matter is the diffusion near rigid walls, that has been extensively studied both theoretically and 
experimentally~\cite{brenner_1961_observation,faucheux_confined_1994,dufresne_2000_hydrodynamic,peters_efficient_2002,felderhof_effect_2005,li_amplified_2008,simonnin_diffusion_2017}. Besides the practical interest of such a configuration towards the sensing of surface forces~\cite{lavaud_stochastic_2021}, one of the key aspects lies perhaps in the apparition of anisotropic and space-dependent mobilities~\cite{matse_2017_test} due to the frictional boundary condition at the walls. As a consequence, multiplicative noises emerge, leading to non-Gaussian diffusion and the probability enhancement of rare events~\cite{wang_anomalous_2009,Barkai2020,alexandre_how_2022, millan_numerical_2023,xu2024}. 
    \begin{figure}[t!]
    \centering
    \includegraphics[width=0.3\textwidth]{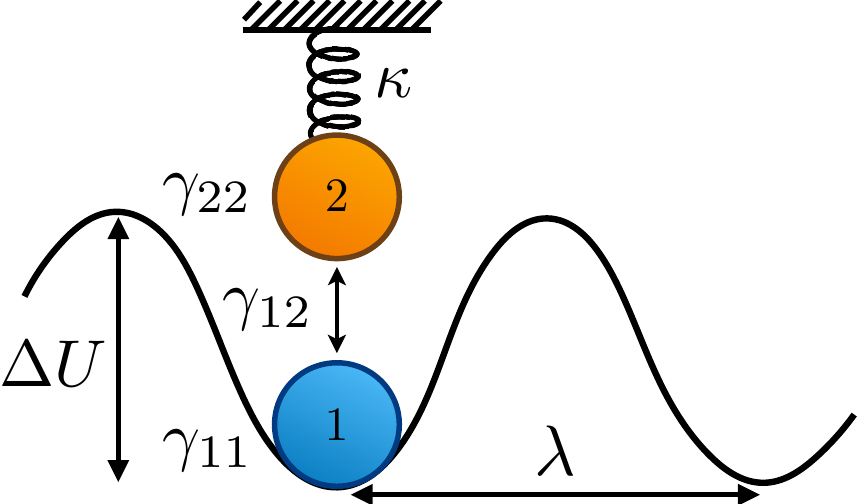}
    \caption{Schematic of the system. 
A test Brownian particle \( (1) \) is trapped in a periodic energy potential of amplitude \(\Delta U\)
         and spatial period \(\lambda\). It is
coupled via a hydrodynamic-like interaction of constant damping coefficient \(\gamma_{12}\) to a thermally-fluctuating 
elastic mode, represented by another Brownian particle \( (2) \) trapped in a harmonic trap of compliance (\textit{i.e.} inverse stiffness) $\kappa$. 
Besides, both particles feel individual hydrodynamic drag forces represented by 
        their constant self-drag coefficients \(\gamma_{11}\) and \(\gamma_{22}\), and are described within the overdamped Langevin regime.}
    \label{fig:fig_1}
\end{figure}
 
However, in most practical situations where molecules or microorganisms diffuse near surfaces, the latter are typically soft. Therefore, the dynamical properties of Brownian particles near deformable boundaries must be understood. Along this line of thought, a few studies, mostly theoretical, have addressed the case of tracers diffusing near soft surfaces. Examples include liquid interfaces~\cite{bickel_brownian_2006,Wang2009}, and fluctuating ones~\cite{marbach_transport_2018}. The case of elastic interfaces was also addressed in the contexts of contactless microrheology~\cite{Bar2017} and biological membranes~\cite{Daddi2018}. Nevertheless, these works focused on the low-coupling regime, where the particles are distant or point-like, and thus unable to generate rectified flows that lead to surface deformation. Interestingly though, within a purely deterministic context, near-contact elastohydrodynamic couplings have been studied in the past decades. Examples include contactless colloidal-probe rheology of soft substrates in normal mode~\cite{Leroy2012,Wang2015}, as well as the emergence of lift forces at low Reynolds numbers~\cite{Bureau2023,Rallabandi_2024_Fluid}, that were first revealed theoretically~\cite{sekimoto1993mechanism,Beaucourt2004,skotheim_soft_2004,urzay_elastohydrodynamic_2007,Snoeijer2013}, and then confirmed experimentally~\cite{Saintyves2016,Davies2018,Rallabandi2018,Vialar2019,zhang2020direct}. The leading-order corrections induced by elasticity to the full mobility tensor near a rigid wall~\cite{brenner_1961_observation} were obtained for various geometries and rheologies~\cite{salez_elastohydrodynamics_2015,Bertin2022,Jha2024}, and revealed a zoology of counter-intuitive inertial-like solutions.

In view of the above state of the art, the arising challenge is now to understand the influence of soft boundaries on the spontaneous motion of nearby finite-sized Brownian particles. Molecular-dynamics simulations were performed in such a context~\cite{Sheikh2023}. Besides, a recent experimental study indicated that soft colloids diffusing near rigid walls feel transient, non-conservative forces that could be related to an intricate combination of thermal fluctuations, soft boundaries and viscous flows~\cite{fares_observation_2024}. 
Finally, at a practical level, a natural question to ask is wether or not the effects 
 of fast and tiny surface fluctuations can be observed over the long times and large distances  associated with colloidal diffusion.
In this Letter, we thus introduce a minimal, but generic, theoretical model in order to investigate Brownian elastohydrodynamic effects. Specifically, we combine the hydrodynamic coupling of a colloidal particle with an elastic mode~\cite{Ber2016, Sal2020} to an external periodic trapping~\cite{Reim2001, Lee2006, Linder2016, Bel2022}. We solve the problem both analytically and numerically, and demonstrate that the late-time diffusion coefficient of the particle
increases notably with the compliance of the elastic mode.

{ \em The model.}\ As depicted in Fig.~\ref{fig:fig_1}, we consider an overdamped system made of two coupled degrees of freedom: i) a principal degree of freedom $q_1(t)$, corresponding to the spatial position of a test colloid over time $t$, which 
is free to explore space while being subject to a periodic trapping potential $\phi(q_1)$; and ii) an auxiliary degree of freedom $q_2(t)$, corresponding to an elastic mode, which is 
harmonically confined with a potential energy $q_2^{\,2}/(2\kappa)$, where $\kappa$ is the compliance. 
Thus, the total energy of the system is given by the Hamiltonian
\begin{equation}
H(q_1,q_2) = \phi(q_1) +\frac{q_2^{\,2}}{2\kappa} ~.
\end{equation}
At thermal equilibrium, the probability density function (PDF) of $q_1$ and $q_2$ is given by the Gibbs-Boltzmann distribution
\begin{equation}
P_\mathrm{eq}(q_1,q_2) = \frac{\exp\left[-\beta H(q_1,q_2)\right]}{Z}~,\label{gbd}
\end{equation}
where $Z$ is the normalising partition function,  and where $\beta =1/(k_\mathrm{B}T)$ with $T$ the temperature 
and $k_\mathrm{B}$ Boltzmann's constant. The equilibrium statistics of $q_1$ and $q_2$ are independent due to the additivity of the Hamiltonian, which implies in particular that $q_2$ does not affect the equilibrium distribution of $q_1$.

The system dynamics is described by the coupled overdamped Langevin equations
\begin{gather}
    \gamma_{11} \frac{\mathrm{d}q_1}{\mathrm{d}t} +\gamma_{12} \frac{\mathrm{d}q_2}{\mathrm{d}t}   = - \phi'(q_1) + \eta_1(t)~,  \label{odlangevin1} \\
    \gamma_{12} \frac{\mathrm{d}q_1}{\mathrm{d}t} +\gamma_{22} \frac{\mathrm{d}q_2}{\mathrm{d}t}  = - \frac{q_2}{\kappa} + \eta_2(t)~,
    \label{odlangevin2}
\end{gather}
where the constants $\gamma_{ij}$, with $i,j \in \{1,2\}$, are the components of the friction tensor $\gamma$, which is related to the mobility tensor $\mu$ through $\mu =\gamma^{-1}$; and where
 the functions $\eta_i (t)$ are zero-mean Gaussian white noises, with a correlation function
 \(\langle \eta_i(t) \eta_j(t') \rangle = 2k_\mathrm{B}T\delta(t-t') \gamma_{ij}\), where $\langle\cdot\rangle$ is the ensemble average.
  This choice of noise correlation is self-consistent with the equilibrium distribution of $q_1$ and $q_2$ being given by Eq.~(\ref{gbd}).
 
{ \em Theoretical analysis.} In the weakly-compliant limit ($\kappa\ll1$), 
one can compute the late-time diffusion coefficient $D^*(\kappa)$ of the colloid using a perturbation analysis at $\cal{O}(\kappa)$, together with an exact representation of the effective diffusion coefficient obtained via the Stokes-Einstein relation, and a multiscale analysis~\cite{hai2008,pav2008}. Doing so (see details in SM~\cite{SupplementaryMaterial}), we obtain
\begin{equation}
    D^*(\kappa) \simeq D^*(0) \left(1 + \alpha\right)~,
    \label{Dean-Lifson-Jackson}
\end{equation}
with
\begin{equation}
    \alpha = \beta \kappa\frac{\gamma_{12}^2}{\gamma_{11}^2} 
    \frac{\int_{0}^{\lambda}\mathrm{d}q_1\, \phi'(q_1)^2\,\mathrm{exp}\left[\beta \phi(q_1)\right] }
    {\int_{0}^{\lambda}\mathrm{d}q_1\, \mathrm{exp}\left[\beta \phi(q_1)\right]}~,
    \label{alpha_factor}
\end{equation}
and
\begin{equation}
D^*(0) = \frac{k_\mathrm{B} T \lambda^2}{\gamma_{11} Z_+Z_-}~,
\end{equation}
where 
\begin{equation}
Z_\pm= \int_{0}^{\lambda}\mathrm{d}q_1\, \mathrm{exp}\left[\pm\beta \phi(q_1)\right]~.
\end{equation}
Since $\alpha\geq0$, the hydrodynamic coupling of the colloid to the elastic mode always increases the late-time diffusion coefficient of the colloid. Moreover, the effect is larger at larger compliance $\kappa$. We also see that a non-flat periodic potential $\phi(q_1)$ is required for the effect to exist, within the assumptions of the model. 

In the particular case of a sinusoidal potential, where
$\phi(q_1) = \Delta U \sin(2\pi q_1/\lambda)$, with $\Delta U$ the amplitude of the potential, and $\lambda$ its spatial period, we have
\begin{equation}
\alpha = 4\pi^2\epsilon\frac{\gamma_{12}^2}{\gamma^2_{11}}\frac{v I_1(v)}{I_0(v)}~,\label{eps}
\end{equation}
with $v= \beta\Delta U$, where $I_n$ denotes the order-$n$ modified Bessel function of the first kind, and where we introduced the dimensionless compliance\begin{equation}
\epsilon= \frac{\kappa k_{\mathrm{B}}T}{\lambda^2}~.\label{deps}
\end{equation}
As such, the relative increase of the late-time diffusion coefficient is larger for smaller $\lambda$. This is attributed to the fact that for a smaller $\lambda$ the coupling to the elastic mode increases the jump rate of the colloid between neighbouring minima, instead of being averaged out as would be the case at large $\lambda$. We also see that the relative increase of the late-time diffusion coefficient depends linearly on the function $vI_1(v)/I_0(v)$,
which is a monotonically increasing function of $v$. This means that the relative magnitude of the effect increases with
 the depth of the trap -- the obvious counterpart being that larger times are then needed to reach the late-time diffusive regime. Finally, the relative magnitude of the effect increases with the relative magnitude $\gamma_{12}/\gamma_{11}$ of the hydrodynamic coupling to the elastic mode, as expected. 
 
The multiscale analysis can also be employed in the limit where $\epsilon\to\infty$, which corresponds to a very compliant elastic mode. In this case, we also find (see details in SM~\cite{SupplementaryMaterial}) that the late-time diffusion coefficient is increased relatively to the rigid-case value, and saturates to a maximum value at large $\epsilon$, as
 \begin{equation}
 D^*\simeq\frac{k_{\textrm{B}} T \mu_{11}(\mu_{11}\mu_{22}-\mu_{12}^2)}{\lambda^{-2}Z_+Z_-(\mu_{11}\mu_{22}-\mu_{12}^2)+\mu_{12}^2}+O\left(\frac{1}{\sqrt{\epsilon}}\right)~.
 \label{soft}
\end{equation}
Such a plateau value can be notably higher than the rigid-case value, indicating the potential strength of the effect in practical systems relevant to soft matter and biophysics. Besides, once again, we note that when the periodic potential $\phi(q_1)$ is flat, the late-time diffusion coefficient is not modified by the hydrodynamic coupling to the elastic mode. 

{ \em Heuristic argument.} In order to understand the softness-induced late-time diffusivity enhancement from a physical point of view, we now present a much simpler heuristic argument giving the same result as the rigorous analysis in the small-$\epsilon$ limit (see details in SM~\cite{SupplementaryMaterial}).
The intrinsic time scale $\tau_2 =\kappa\gamma_{22}$ of the stochastic process $q_2(t)$ is assumed to be small with respect to all the other time scales. In this limit, the effective zero-temperature overdamped Langevin equation for \(q_1(t)\) is
 \begin{equation}
    \frac{\mathrm{d}q_1}{\mathrm{d}t} = -\mu_\mathrm{e}(q_1) \phi'(q_1)~,
    \label{zerotempq1}
\end{equation}
where we have introduced the dressed mobility \begin{equation}
    \mu_\mathrm{e}(q_1) = \frac{1}{\gamma_{11}\left[1 +\frac{\kappa\gamma_{12}^2\phi''(q_1)}{\gamma_{11}^2}\right]}~.
    \label{mueffective}
\end{equation}

The equilibrium marginal PDF in position $q_1$ is $P_{\textrm{meq}}(q_1)=(Z_-)^{-1}\exp[-\beta \phi(q_1)]$. Therefore, assuming that the time-dependent marginal PDF $P_1(q_1,t)$ satisfies an effective Fokker-Planck equation, the latter must have the form
\small
\begin{equation}
    \frac{\partial P_1(q_1,t)}{\partial t} = \frac{\partial}{\partial q_1}\left\{\mu_\mathrm{e}(q_1) 
    \left[k_\mathrm{B}T\frac{\partial P_1(q_1,t)}{\partial q_1}+\phi'(q_1)P_1(q_1,t)\right]\right\}~,
    \label{fokker-planck}
\end{equation}
\normalsize
where the $T$-independent term on the right-hand side is determined from the zero-temperature dynamics, while the $T$-dependent term on the right-hand side is constructed to ensure convergence of $P_1$ to $P_{\textrm{meq}}$ at infinite time. Note that we have implicitly assumed here that the dressed mobility $\mu_\mathrm{e}(q_1)$ 
 does not exhibit any temperature dependence.

The dressed diffusion coefficient is given by the Stokes-Einstein relation,  \(D_\mathrm{e}(q_1) = k_\mathrm{B}T\mu_\mathrm{e}(q_1)\), 
which differs from the bare (or ``microscopic") diffusion coefficient $D_\mathrm{m}=k_\mathrm{B}T/\gamma_{11}$ of the colloid. Interestingly, the spatial dependence of the dressed diffusion coefficient implies that 
the effective Langevin equation must have a multiplicative noise, leading to a supplementary source of non-Gaussian 
displacements~\cite{wang_anomalous_2009,Barkai2020,lavaud_stochastic_2021,alexandre_how_2022, millan_numerical_2023,xu2024, sposini2024} in addition to the trivial one induced by the trapping potential. 
 
We now invoke the Lifson-Jackson 
formula~\cite{lifson_self-diffusion_1962}, providing the late-time diffusion coefficient in a periodic potential, which reads 
\begin{equation}
    D^*(\kappa) = \frac{ \lambda^2}{\int_0^\lambda \mathrm{d}q_1  \frac{\mathrm{exp}\left[\beta\phi(q_1)\right]}{D_{\textrm{e}}(q_1)} 
     \int_0^\lambda \mathrm{d}q_1 \mathrm{exp}\left[-\beta\phi(q_1)\right]}~.
    \label{lj_applied}
\end{equation}
From this expression, we can recover Eqs.~(\ref{Dean-Lifson-Jackson}) and~(\ref{alpha_factor}). We emphasize that 
 the heuristic argument we presented here is not rigorous. Nevertheless, it can also be applied to evaluate the effective diffusion coefficient
of an underdamped Brownian particle near the overdamped limit (see details in SM~\cite{SupplementaryMaterial}), for which we once again recover the results obtained via rigorous multiscale analysis \cite{pav2008,hai2008}. The same
   method also yields the first-order correction to the effective diffusion coefficient for a particle carrying a dipole moment which 
   interacts with a random or periodic electric field when the dipole dynamics is assumed to be much faster than the particle diffusion~\cite{tou2009}.
   
{ \em Numerical simulations.} In order to verify the above predictions, Eqs.~(\ref{odlangevin1}) and~(\ref{odlangevin2}) are numerically 
solved for a sinusoidal trapping potential $\phi(q_1)$, using the Euler-Maruyama scheme~\cite{maruyama_continuous_1955}, through an
optimised \emph{Cython}~\cite{behnel2010cython} code (see details in SM~\cite{SupplementaryMaterial}). The initial positions of the test particle and the elastic mode are sampled from Eq.~(\ref{gbd}).
\begin{figure}[t!]
    \hspace{-.2in}
  \includegraphics[width=0.5\textwidth]{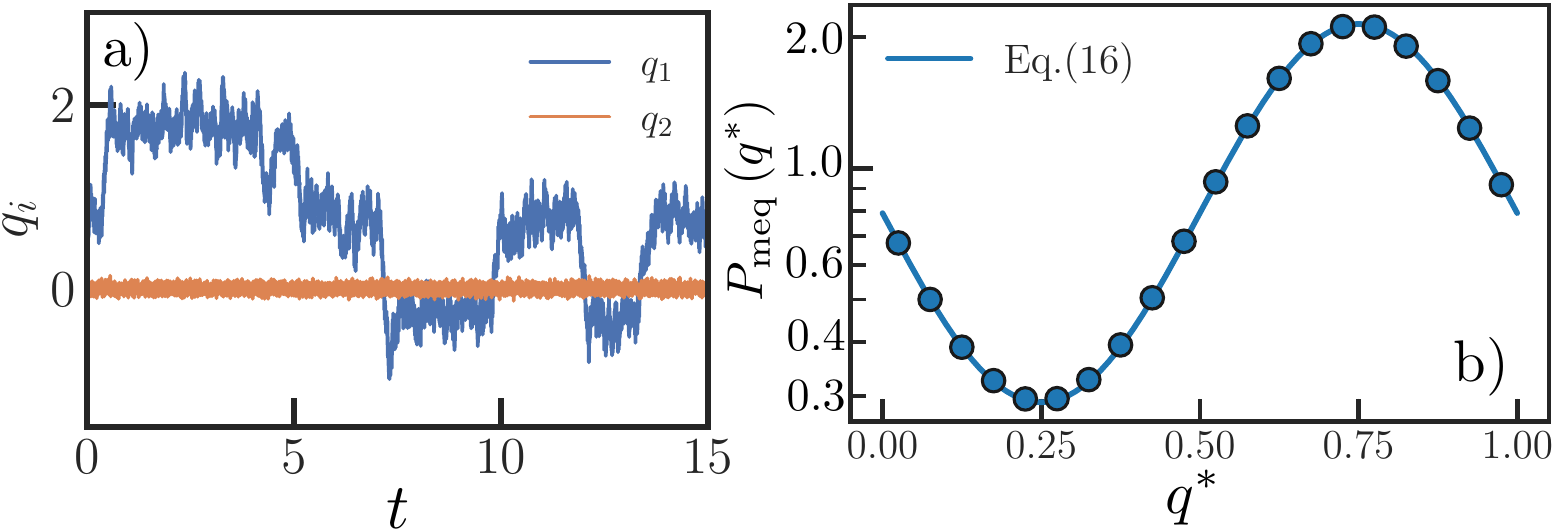}
    \caption{a) Typical trajectories $q_i(t)$ of the test particle ($i=1$, blue line) and the elastic mode ($i=2$, orange line), 
    simulated by numerically solving Eqs.~(\ref{odlangevin1}) and~(\ref{odlangevin2}), 
    for \(\mu _{11}\)= 1.0, \(\mu_{12}\)= 0.5, \(\mu _{22}\) = 0.5,
    \(\Delta U\) = 1.0, \(\lambda\) = 1.0, \(\kappa = 1.0\times 10^{-3}\), \(\Delta t = 1.0\times10^{-4}
    \), and \(k_\mathrm{B}T\) = 1.0.
    b) Late-time marginal PDF  \(P_\mathrm{meq}\) for the periodised 
    position $q^*$, given by $q_1= q^*{\rm mod}(\lambda)$, of the test particle,
     obtained by binning \(10^2\) trajectories of \(10^7\) points. The symbols are the results from the numerical simulations. The solid blue line corresponds to the Boltzmann 
     distribution given in 
     Eq.~(\ref{Peqp}). 
   \label{fig:fig2}}
\end{figure}
Typical raw trajectories are shown in Fig.~\ref{fig:fig2}a). We observe that the elastic mode $q_2(t)$ fluctuates around its equilibrium value (\(q_2 = 0\)), 
within the harmonic potential
  it is subject to, whereas the position $q_1(t)$ of the test colloid fluctuates at short times in a minimum
   of the sinusoidal potential $\phi(q_1)$ before randomly jumping from one minimum to the other by passing the potential barrier. From such trajectories, we can first compute the late-time PDF in positions. As an example, 
   Fig.~\ref{fig:fig2}b) shows the equilibrium marginal PDF of the periodised variable $q^*$, defined by $q_1 = q^*{\rm mod}(\lambda)$, which satisfies
\begin{equation}
P_\mathrm{meq}(q^*) = \frac{\exp[-\beta \phi(q^*)]}{\int_0^\lambda \mathrm{d}q\exp[-\beta \phi(q)]}~.\label{Peqp}
 \end{equation}  
\begin{figure}[t!]
    \includegraphics[width=0.45\textwidth]{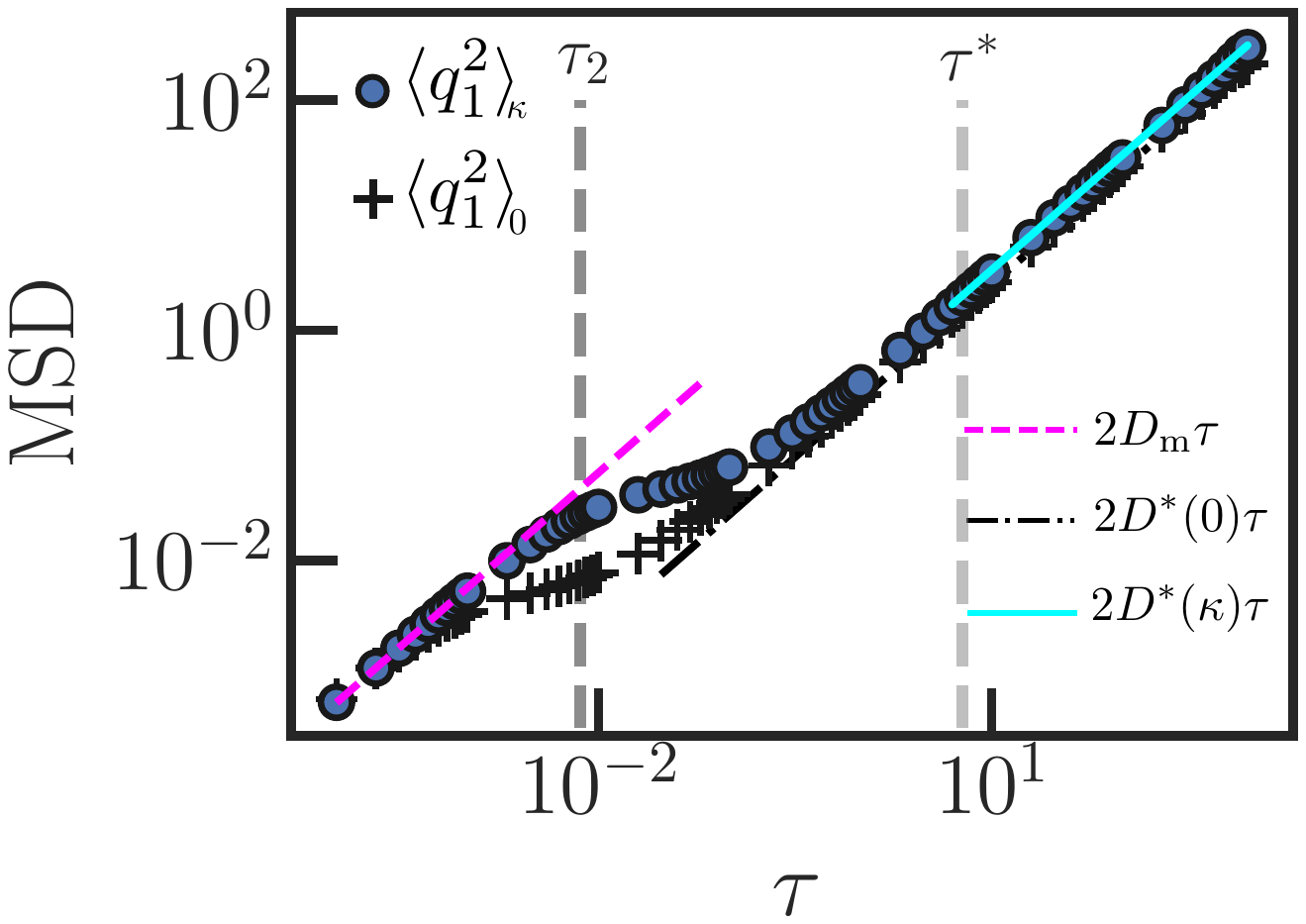}
    \caption{Mean-squared displacements \( \langle q_1 ^2 \rangle_{\kappa}\) and  \( \langle q_1 ^2 \rangle_{0}\)
    of the test particle as functions of the time increment $\tau$, for \(\kappa = 1.0\times 10^{-3}\)
     (blue disks) and \(\kappa = 0.0\) (black crosses) respectively, 
    as computed from 
   \(10^2\) trajectories of \(10^7\) points each, for \(\gamma _{11}\)= 5.0, \(\gamma_{12}\)= 22.0, \(\gamma_{22}\) = 104.0,
    \(\Delta U\) = 1.0, \(\lambda\) = 1.0, \(\Delta t = 1.0\times10^{-4}
    \), and \(k_\mathrm{B}T\) = 1.0. The symbols are the results from the numerical simulations. The dash-dotted black line indicates $2D^*(0)\tau$, where $D^*(0)$ is the classical Lifson-Jackson late-time diffusion 
    coefficient~\cite{lifson_self-diffusion_1962}. The solid cyan line indicates $2D^*(\kappa)\tau$, where $D^*(\kappa)$ is the late-time diffusion coefficient given by 
    Eq.~(\ref{Dean-Lifson-Jackson}). 
    The dashed pink line indicates $2 D_\mathrm{m}\tau$, where $D_\mathrm{m} = k_\mathrm{B}T/ \gamma_\mathrm{11}$ is the bare diffusion coefficient. The characteristic times \( \tau_2 = \kappa\gamma_{22}\) and \( \tau^*  = \lambda^2 / D^*(\kappa) \) are indicated with the dark-grey and light-grey vertical dashed lines,
    respectively.}
      \label{fig:fig3} 
\end{figure}

In order to probe the dynamical effects of the hydrodynamic coupling of the trapped test particle to the elastic mode, 
we now turn to the analysis of the mean-squared displacement (MSD), defined as
\begin{equation}
    \langle q_i^2 \rangle (\tau) = \langle [q_i(t+\tau) - q_i(t)]^2\rangle~,
\end{equation}
where $\tau$ is a time increment. As shown in Fig.~\ref{fig:fig3}, the MSD of the test colloidal particle exhibits two distinct diffusive regimes separated 
by a crossover region. The short-time diffusive regime is characterized by the bare diffusion coefficient $D_{\textrm{m}}$. The late-time diffusive regime is characterized by the late-time diffusion coefficient $D^*$, and is reached when the particle has diffused over many periods of the trapping potential, i.e. when its MSD is large compared to $\lambda^2$. This sets a time scale
$\tau^* = \lambda^2/D^*$, which must be exceeded to see the late-time diffusive regime.

\begin{figure}[t!]
    \hspace*{-0.3in}
    \includegraphics[width=0.4\textwidth]{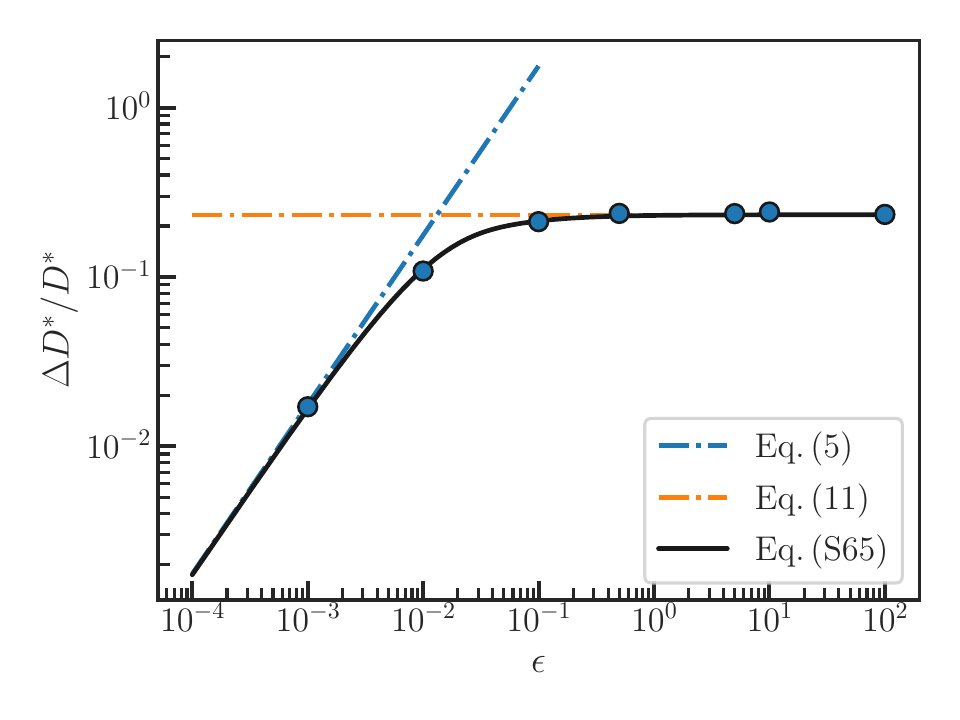}
    \caption{Relative increase $\Delta D^* / D^*=[D^*(\epsilon)-D^*(0)] / D^*(0)$ of the late-time diffusion coefficient $D^*$ 
    of the test particle as a function of the dimensionless compliance $\epsilon$, for \(\mu _{11}\)= 1.0, \(\mu_{12}\)= 0.5, \(\mu _{22}\) = 0.5,
    \(\Delta U\) = 1.0, \(\lambda\) = 1.0, and \(k_\mathrm{B}T\) = 1.0. The symbols are the results from the numerical simulations. The dash-dotted blue
      line corresponds to the small-$\epsilon$ prediction of Eq.~(\ref{Dean-Lifson-Jackson}). The dash-dotted orange
      line corresponds to the large-$\epsilon$ prediction of Eq.~(\ref{soft}). The solid black line
      corresponds to the Kubo formula of Eq.~(S65) (see details in SM~\cite{SupplementaryMaterial}).}
       \label{fig:fig4}
\end{figure}

Finally, we focus on the late-time diffusive regime of the simulated trajectories, and compute from the MSD the relative excess in late-time diffusion coefficient, i.e. $\Delta D^* / D^*=[D^*(\epsilon)-D^*(0)] / D^*(0)$, as a function of
   $\epsilon$. The results are shown in Fig.~\ref{fig:fig4} for a given set of parameters. We observe that the numerical-simulation data agree well with the small-$\epsilon$ and large-$\epsilon$ asymptotic predictions of Eqs.~(\ref{Dean-Lifson-Jackson}) and~(\ref{soft}), respectively. Furthermore, the Kubo formula can be used to evaluate $D^*$ for any value of $\epsilon$ (see details in SM~\cite{SupplementaryMaterial}, and in particular Eq.~(S65)). We see that the numerical-simulation data agree at all $\epsilon$ values with the Kubo formula, and that the latter interpolates well the two asymptotic predictions. 

In conclusion, we theoretically and numerically studied the Brownian motion of colloid-like particles near elastic interfaces using minimal ingredients. 
We showed that a Brownian particle trapped in a periodic potential sees its late-time diffusivity
dressed and significantly enhanced by hydrodynamic coupling to a hidden elastic mode. Despite being minimalist, our description reveals generic, 
 original and large effects, relevant for the transport of colloidal particles near complex boundaries. In particular, our results suggest the intriguing possibility of understanding, and even tuning, the diffusivity of colloids near deformable and patchy interfaces -- with key implications for biophysics and nanofluidics.
\newline

\emph{Data availability. } All codes used in this study to produce
data are available through the online repository \url{https://github.com/EMetBrown-Lab/Lacherez2024\textunderscore Codes}.
\newline

\emph{Acknowledgements. } The authors thank Nicolas Fares, Quentin Ferreira, Aditya Jha and Elodie Millan for interesting discussions. 
Computer time for this study was provided by the computing facilities of the M\'esocentre de Calcul Intensif Aquitain.
 The authors acknowledge financial support from the European Union through the European Research Council under EMetBrown 
 (ERC-CoG-101039103) grant. Views and opinions expressed are however those of the authors only and do not necessarily reflect 
 those of the European Union or the European Research Council. Neither the European Union nor the granting authority can be held 
 responsible for them. The authors also acknowledge financial support from the Agence Nationale de la Recherche under
  EDIPS (ANR-23-CE30- 0020), Softer (ANR21-CE06-0029) and Fricolas (ANR-21-CE06-0039) grants, as well as from the Interdisciplinary and 
 Exploratory Research Program under a MISTIC grant at the University of Bordeaux, France. They acknowledge as well the support from the LIGHT
 S{\&}T Graduate Program (PIA3 Investment for the Future Program, ANR-17EURE-0027). Finally, they thank the RRI Frontiers of Life, which 
  received financial support from the French government in the framework of the University of Bordeaux's France 2030 program,
   as well as the Soft Matter Collaborative Research Unit, Frontier Research Center for Advanced Material and Life Science, 
 Faculty of Advanced Life Science, Hokkaido University, Sapporo, Japan, and the CNRS International Research Network between France and India on ``Hydrodynamics at small scales: from soft matter to bioengineering".
\bibliographystyle{unsrt}
\bibliography{Lacherez2025}
\end{document}

\title{Enhanced diffusion over a periodic trap by hydrodynamic coupling to an elastic mode \\ -- \textit{Supplementary Material} --}
\author{Juliette Lacherez$^1$}
\author{Maxime Lavaud$^{1,2}$}
\author{Yacine Amarouchene$^1$}
\email{yacine.amarouchene@u-bordeaux.fr}
\author{David S. Dean$^1$}
\email{david.dean@u-bordeaux.fr}
\author{Thomas Salez$^1$}
\email{thomas.salez@cnrs.fr}
\affiliation{$^1$Univ. Bordeaux, CNRS, LOMA, UMR 5798, F-33400, Talence, France}
\affiliation{$^2$Univ. Bordeaux, CNRS, Bordeaux INP, CBMN, UMR 5248, F-33600, Pessac, France}
\date{\today}
\maketitle

We consider and over over-damped system made up of two degrees of freedom, the principal degree of freedom $q_1$ which 
is free to explore a large region of space, for example it is subject to a periodic potential $\phi(q_1)$. The second degree of freedom $q_2$ is a highly constrained one and is taken to be harmonically confined with potential energy $\frac{1}{2\kappa} q_2^2$, where we will assume that $\kappa$ the spring compliance is small. The total energy of the system is thus given by the Hamiltonian
\begin{equation}
H(q_1,q_2) = \phi(q_1) +\frac{1}{2\kappa} q_2^2~.
\end{equation}
In thermal equilibrium the steady state of probability distribution of $q_1$ and $q_2$  is given by the Gibbs-Boltzmann distribution
\begin{equation}
P_{eq}(q_1,q_2) = \frac{\exp[-\beta H(q_1,q_2)]}{Z}~,\label{gbd}
\end{equation}
where $Z$ is the normalising partition function and $\beta =1/(k_{\mathrm{B}}T)$ is the inverse thermal energy, where $T$ is the temperature and  $k_{\mathrm{B}}$ is  Boltzmann's constant. In equilibrium, the statistics of $q_1$ and $q_2$ are independent due to the additivity of the Hamiltonian $H$, {\em i.e.} $P_{eq}(q_1,q_2)=P_{1eq}(q_1)P_{2eq}(q_2)$. However dynamically they can be coupled; for instance this model can be seen as a toy model for a colloidal particle near an elastic compliant surface. As such, any single time observations  in equilibrium of the measurable quantity $q_1$ will be independent of the effect of $q_2$ and hence will not show any effect due to the hydrodynamic coupling with $q_2$. This means that one would have to measure at least two time correlation functions but due to the rapid dynamics of $q_2$ one would have to resolve the dynamics on very short time scales to detect the hydrodynamic coupling. 

The simplest form for the over-damped Langevin dynamics of this system is 
\begin{eqnarray}
\frac{dq_1}{dt} &=& -\mu_{11} \phi'(q_1) - \frac{\mu_{12}}{\kappa} q_2 + \eta_1(t)\label{adq1}\\
\frac{dq_2}{dt} &=& -\frac{\mu_{22}}{\kappa} q_2  -\mu_{12} \phi'(q_1) + \eta_2(t)~,\label{adq2}
\end{eqnarray}
where the mobility coefficients $\mu_{ij}$ are taken for simplicity to be constant, and the noise correlation function is chosen to ensure that this Langevin system has an steady state  distribution given by Eq.~ (\ref{gbd}), explicitly
\begin{equation}
\langle \eta_{i}(t)\eta_j(t)\rangle = 2k_{\mathrm{B}}T \delta(t-t') \mu_{ij}~.
\end{equation}
From a mathematical point of view, the constraints on the mobility matrix are that it is symmetric and has positive eigenvalues. However from a physical point of view, in the absence of coupling we must have that
\begin{equation}
\mu_{11} > 0, {\rm and }\ \mu_{22}~.
\end{equation}
The constraint on the eigenvalues of the mobility matrix then translates to
\begin{equation}
\det(\mu) = \mu_{11} \mu_{22}-\mu_{12}^2 >0~.
\end{equation}
In an experimental context it may be possible to directly observe
the motion of the variable $q_1$, but not necessarily on very short time scales, while the direct observation  of the second variable $q_2$ may be impossible because it cannot be resolved spatially or temporally. If the variable $q_1$ is subject to a confining potential and a  measurement is made of its equilibrium statistics, the variable $q_1$ has no effect and changing the value of $\kappa$ for instance will not affect the measurements. Observing the dynamics of the variable $q_1$ may yield some information about the variable $q_2$ but if the time scale over which $q_2$ equilibrates is too fast, again this information is censored by being averaged out.

Here we propose a solution to the demonstrate the effect of the variable $q_2$ by examining its effect on the late time diffusion constant of $q_1$ when it is subjected to a periodic potential of the type that can be generated by an optical trap. Given that we are examining diffusion at large times and over large distances, both temporal and spatial resolution of the experiment are not an issue. The effect of the variable $q_2$ is determined by its influence on how $q_1$ moves in the potential $\phi$ and how this influence integrates overtime to change the diffusion constant of $q_1$. We will see that there is an intrinsic time scale associated with the dynamics of the elastic variable $q_2$ and we assume that this time scale is small and thus difficult to resolve. Measurements of $q_1$ made over time scales larger than $\tau_2$ will not see the effect of $q_2$ due to averaging. If the variable $q_1$ diffuses in a periodic potential and the typical time to cross the potential barriers (so to jump from minimum to minimum) is of the order of $\tau_2$ then the first passage time from minimum to minimum will be affected by the dynamics of $q_2$ and thus show up in the late time diffusion constant of $q_1$ thus yielding a mechanism for revealing the hydrodynamic coupling. Rather than trying to resolve short time scales directly we are thus imposing a short distance, and this can be done with an optical trap. 
\section{Heuristic argument for the correction to the late time diffusion constant \label{heur}}

In order to determine the effective Langevin dynamics for the variable $q_1$ in the limit where $\kappa$ is small, and so the variable $q_2$ is small, we consider the dynamics of the system without any noise. Using the subsequently derived effective dynamics for $q_1$ at zero temperature, we then add white noise to the system in order to ensure that the resulting Langevin equation has the correct marginal equilibrium probability distribution
\begin{equation}
P_{meq}(q_1) = \frac{\exp[-\beta \phi(q_1)]}{Z_-}~.
\end{equation}
where we use the notation
\begin{equation}
Z_\pm=\int_0^\lambda dq \exp[\pm\beta \phi(q)]~,
\end{equation}
where $\lambda$ is the period of the potential.
This is clearly not a rigorous procedure and we will justify the result rigorously below using multiscale analysis. However the it should be noticed that this method can also be applied to  compute the correction to diffusion of a Brownian particle  in a periodic potential due to
inertia (the deviation from the strongly over-damped limit) \cite{pav2008,hai2008} and actually recovers first order correction (for completeness we give a derivation later in this SM).  It can also be applied to compute the correction to the diffusion constant of a Brownian particle carrying an electric dipole in a periodic electric field and again the simple analysis agrees with a more rigorous one using stochastic analysis \cite{tou2009}.

We proceed by formally solving Eq.~ (\ref{adq2}) at $T=0$ (that is to say without noise) to give
\begin{equation}
q_2 =-\frac{\kappa\mu_{12}}{\mu_{22}} (1 +\frac{\kappa}{\mu_{22}}\frac{d}{dt})^{-1}\phi'(q_1)
\end{equation}
where $\tau_2 =\frac{\kappa}{\mu_{22}}$ is a short time scale due to the fact that $\kappa$ is small. Carrying out an operator expansion for small $\tau_2$ we find
\begin{equation}
q_2 \simeq-\frac{\kappa\mu_{12}}{\mu_{22}} (\phi'(q_1) -\frac{\kappa}{\mu_{22}}\phi''(q_1)\frac{dq_1}{dt} + O(\tau_2^2)).
\end{equation}
Now substituting this in Eq.~ (\ref{adq1}) gives
\begin{equation}
\frac{dq_1}{dt} = -(\mu_{11}-\frac{\mu_{12}^2}{\mu_{22}})\phi'(q_1(t))-\frac{\kappa\mu_{12}^2}{\mu^2_{22}}\phi''(q_1(t))\frac{dq_1}{dt}
\end{equation}
and from this 
\begin{equation}
\frac{dq_1}{dt} = -\mu_{e}(q_1)\phi'(q_1)\label{eq1zt}
\end{equation}
where $\mu_{e}(q_1)$ is the effective mobility 
\begin{equation}
\mu_e(q_1) = \frac{\mu_{11}\mu_{22} -\mu_{12}^2}{\mu_{22}(1 +\frac{\kappa\phi''(q_1)\mu_{12}^2}{\mu_{22}^2})}
\end{equation}
In terms of the  friction matrix $\gamma = \mu^{-1}$, the effective mobility can be written in a  simpler form  as follows:
\begin{equation}
\mu_e(q_1) = \frac{1}{\gamma_{11}(1 +\frac{\kappa\phi''(q_1)\gamma_{12}^2}{\gamma_{11}^2})}~.\label{mue2}
\end{equation}
Rather than discuss the noise necessary to render this into a Langevin equation, we note that to have the correct equilibrium distribution the corresponding Fokker-Planck equation must read
\begin{equation}
\frac{\partial P_1(q_1,t)}{\partial t} = \frac{\partial}{\partial q_1}\mu_e(q_1) \left[k_{\mathrm{B}}T\frac{\partial P_1(q_1,t)}{\partial q_1}+\phi'(q_1)P_1(q_1,t)\right]~,\label{fp1}
\end{equation}
and we see that the effective diffusion constant is given by $D_{e}(q_1) = k_{\mathrm{B}} T\mu(q_1)$. Note that in doing this we have added multiplicative noise and a temperature dependent drift if we use the Ito convention of stochastic calculus; however we avoid this discussion by working directly with the Fokker-Planck equation. Notice that when $T=0$, Eq.~ (\ref{fp1}) is the transport equation for Eq.~ (\ref{eq1zt}).

Now, if the potential is periodic with period $\lambda$, we can use the well known Lifson-Jackson formula for the late time effective diffusion constant for Langevin processes in a potential and with varying diffusivity, both of period $\lambda$:
\begin{equation}
D^* = \frac{k_{\mathrm{B}}T\lambda^2}{\int_0^\lambda dq  \frac{\exp[\beta\phi(q)]}{D_e(q)} \int_0^\lambda dq \exp[-\beta\phi(q)]}~.
\end{equation}
In this case we, can easily show, by integration by parts, that
\begin{equation}
D^*(\kappa) = \frac{k_{\mathrm{B}}T\lambda^2}{\gamma_{11} \int_0^\lambda dq \exp[-\beta\phi(q)]\int_0^\lambda dq  \exp[\beta\phi(q)]\left(1-\frac{\kappa\gamma_{12}^2\phi'^2(q)}{k_{\mathrm{B}}T \gamma^2_{11}}\right)}~.
\end{equation}
We thus see that in the regime where $\kappa$ is small the effect of a finite value of $\kappa$ is to increase the long time diffusivity of the particle. We can also write $\phi(q)= \Delta U\psi(\frac{q}{\lambda})$, where $\Delta U$ is an adjustable energy scale (for instance it could be related to the  laser intensity and in which case $\lambda$  would be a wavelength). This  allows us to write
\begin{equation}
D^*(\kappa) = \frac{k_{\mathrm{B}}T}{\gamma_{11} \int_0^1 dp \exp[-\beta\Delta U \psi(p)]\int_0^1 dp  \exp[\beta\Delta U \psi(p)][1-\frac{\kappa\Delta U^2 \psi'^2(p)\gamma_{12}^2}{T  \lambda^2\gamma^2_{11}}]}~.
\end{equation}
Therefore, even if $\kappa$ is very small, increasing the value of $\Delta U$ or decreasing the value of $L$ can be used to make the change in the long time diffusion constant measurable. 

In the absence of any compliance, $\kappa=0$, and no potential we note that
\begin{equation}
D^*(0) = \frac{k_{\mathrm{B}}T}{\gamma_{11}}~,
\end{equation}
this corresponds to the modification of the diffusivity of $q_1$ when the {\em surface} $q_1$ is rigid but still exerts a hydrodynamic interaction.  

From a physical standpoint, the fluctuations of $q_2$ can only influence the dynamics of $q_1$ on the time scale $\tau_2$ before they are averaged out. However, if the typical time that $q_1$ takes to climb over the potential barrier to the next period is of order $\tau_2$, then $q_2$ can effect the long time diffusion of $q_1$. In this way fast, and small degrees of freedom coupled to a slower one can be detected at even at  low temporal resolution provided one has sufficiently good statistics.
By changing variables, we obtain
\begin{equation}
\frac{D^*(\kappa)}{D^*(0)} = \frac{1}{\left( 1-\frac{\int_0^1 dp  \exp[\beta\Delta U\psi(p)]\frac{\kappa\Delta U^2 \psi'^2(p)\gamma_{12}^2}{k_{\mathrm{B}}T  \lambda^2\gamma^2_{11}}}{\int_0^1 dp  \exp[\beta\Delta U \psi(p)]}\right)}~.
\end{equation}
Now,  setting
\begin{equation}
v=\beta\Delta U~,
\end{equation}
gives
\begin{equation}
\frac{D^*(\kappa)}{D^*(0)} = \frac{1}{( 1-\frac{\kappa k_{\mathrm{B}} Tv^2 \gamma_{12}^2}{  \lambda^2\gamma^2_{11}}\frac{\int_0^1 dp  \exp[v\psi(p)] \psi'^2(p)}{\int_0^1 dp  \exp[v\psi(p)]})}~.
\end{equation}
If we take
\begin{equation}
\psi(u) = \sin(2\pi u)~,
\end{equation}
we find 
\begin{equation}
\frac{D^*(\kappa)}{D^*(0)} = \frac{1}{( 1-\frac{4\pi^2\kappa k_{\mathrm{B}} T \gamma_{12}^2}{  \lambda^2 \gamma^2_{11}}\frac{v I_1(v)}{I_0(v)})}~,
\end{equation}
where $I_n(z)$ denotes the modified Bessel function of the first kind.

This can be written as
\begin{equation}
\frac{D^*(\kappa)}{D^*(0)} \approx ( 1+ \alpha)~,
\end{equation}
where
\begin{equation}
\alpha = \frac{4\pi^2\kappa k_{\mathrm{B}}T\gamma_{12}^2}{  \lambda^2\gamma^2_{11}}\frac{v I_1(v)}{I_0(v)}~.
\end{equation}
In the next section we will see that it is natural to write
\begin{equation}
\alpha = 4\pi^2\epsilon\frac{\gamma_{12}^2}{\gamma^2_{11}}\frac{v I_1(v)}{I_0(v)}~,
\end{equation}
where 
\begin{equation}
\epsilon= \frac{\kappa k_{\mathrm{B}}T}{\lambda^2}~,\label{deps}
\end{equation}
is the effective dimensionless parameter that must be small for the above analysis to be valid.

The enhancement of the relative diffusion constant is clearly maximised by choosing small values of $\lambda$ so that the variable $q_2$ can influence the jumping rate between neighbouring minima rather than being averaged out as would be the case if $\lambda$ were large.  We also see that the relative change depends on the function of the energy scale of the periodic potential
\begin{equation}
g(v) = \frac{v I_1(v)}{I_0(v)}~,
\end{equation}
which is a monotonically increasing function of $v$, this means that the relative change is increased on increasing the energy scale $\Delta U$ of the trap. 

The Fokker-Planck equation  in  Eq.~ (\ref{fp1}) corresponds to the Ito stochastic differential equation
\begin{equation}
\frac{dq_1}{dt}= -\mu_e(q_1)\phi'(q_1) + T\mu'_e(q_1) +\sqrt{2T \mu_e(q_1)}\eta(t)~,\label{ito}
\end{equation}
where $\eta(t)$ is zero mean white noise with correlation function $\langle \eta(t)\eta(t')\rangle= \delta(t-t')$. In the next section we will see that this process has exactly the same diffusion constant as the process $q_1$ in the original hydrodynamically-coupled equations (\ref{dq1},\ref{dq2}). Clearly Eq.~ (\ref{ito}) represents a temporal coarse graining of the original process $q_1$ on time scales greater than $\tau_2$. 
Finally at very short time scales $t < \tau_2$ the microscopic time diffusion constant can similarly be deduced from Eq.~ (\ref{dq1}) and is given by
\begin{equation}
D_m = k_{\mathrm{B}}T\mu_{11}= \frac{k_{\mathrm{B}} T\gamma_{22}}{\gamma_{11}\gamma_{22} -\gamma_{12}^2}~.\label{Dm}
\end{equation}
The expression $D_m$ is simply the diffusion constant for the tracer in the absence of any hydrodynamic interactions with $q_2$ and no periodic potential; or, said otherwise, the bulk diffusivity of $q_1$.
\section{Correction to diffusion based on exact formula for diffusivity and homogenisation theory}
For small $\kappa$, $q_2$ is dominated by its fluctuations which scale via equipartition of energy as
 \begin{equation}
 \delta q_2 \sim \sqrt{\kappa k_{\mathrm{B}}  T}
 \end{equation}
Therefore, from now on we will use the dimensionless position variables 
\begin{equation}
q_1= \lambda q\ {\rm and}\ q_2 = \sqrt{\kappa k_{\mathrm{B}}T}u,
\end{equation}
and again we also write $\beta \phi(q_1)= v\psi(\frac{q_1}{\lambda})$.
The coupled Langevin equations can be rewritten in terms of the dimensionless time
\begin{equation}
s = \frac{\mu_{11} k_{\mathrm{B}}T }{\lambda^2}t
\end{equation}
to yield 
\begin{eqnarray}
\frac{dq}{ds} = -v \psi'(q) -\frac{\mu_{12}}{\mu_{11}\sqrt{\epsilon}}u + \eta'_1(s)\\
\frac{du}{ds} = -\frac{\mu_{22}}{\mu_{11}}\frac{u}{\epsilon}-v\frac{\mu_{12}}{\mu_{11}\sqrt{\epsilon}} \psi'(q)+ \frac{1}{\sqrt{\epsilon}}\eta'_2(s)~,
\end{eqnarray}
where $\eta_1'(s)$ and $\eta_2'(s)$ are zero mean Gaussian white noises with correlation functions
\begin{equation}
\langle \eta'_i(s)\eta'_j(s')\rangle = 2 \frac{\mu_{ij}}{\mu_{11}}\delta(s-s')~.
\end{equation}
and $\epsilon$ is that defined in Eq.~ (\ref{deps}).

The late time diffusion coefficient for $q_1$ is then given by via
\begin{equation}
2D^*t \simeq \langle q_1^2(t)\rangle = \lambda^2 \langle q^2(s)\rangle \simeq 2\lambda^2  D_{q_1}^*s = 2 D_{q}\mu_{11} k_{\mathrm{B}}T t~.
\end{equation}
This means that the late time diffusion constant for $q_1$, $D^*$, is related to the late time  diffusion constant of the scaled process $q$ by
\begin{equation}
D^* = \mu_{11} k_{\mathrm{B}}T D^*_q(\epsilon, v, \frac{\mu_{22}}{\mu_{11}}, \frac{\mu_{12}}{\mu_{11}})~,\label{dqd}
\end{equation}
where we have given the explicit functional dependencies for $D^*_q$.
We now carry out a multi-scale   analysis along the lines of \cite{pav2008,hai2008}. where the correction of diffusion of a Brownian particle  in a periodic potential due to
inertial effects was given.  Here , we will use the Stokes-Einstein relation to compute the diffusion constant and so we will add an additional force $h$ acting on the variable $q$ in addition to the periodic potential. 
The Fokker-Planck equation for the variables $q$ and $u$ is then given by
\begin{equation}
 \frac{\partial P}{\partial s} = HP~,\label{fps}
\end{equation}
where $H$ is given by
\begin{equation}
HP= \left(H_0 + \frac{1}{\sqrt{\epsilon}}H_1 + \frac{1}{\epsilon}H_2\right)P~,
\end{equation}
and the corresponding terms are given by
\begin{eqnarray}
H_0P&=& \frac{\partial}{\partial q}\left(\frac{\partial P}{\partial q} + [v\psi'(q)- h]P \right)\\
H_1P&=&\frac{\mu_{12}}{\mu_{11}}\left[u\frac{\partial P}{\partial q} + 2\frac{\partial^2P
}{\partial q \partial u} +(v\psi_1'(q)- h) \frac{\partial P }{\partial u}\right]\\
H_2P&=& \frac{\mu_{22}}{\mu_{11}}\frac{\partial}{\partial u}\left( \frac{\partial P}{\partial u}+u P\right)~.
\end{eqnarray}
The expansion we will make is based on small $\epsilon$ but also depends on the parameters $\mu_{22}/\mu_{11}$ and
$\mu_{12}/\mu_{11}$, the separation of scales needed to implement the expansion is more precisely states as
\begin{equation}
1\ll \frac{1}{\sqrt{\epsilon}}\frac{\mu_{12}}{\mu_{11}}\ll \frac{1}{\epsilon}\frac{\mu_{22}}{\mu_{11}}~,
\end{equation}
and notably the second inequality implies that 
\begin{equation}
\frac{\mu_{12}}{\mu_{11}}\ll \frac{1}{\sqrt{\epsilon}}\frac{\mu_{22}}{\mu_{11}}~.
\end{equation}
It will be  useful later to employ the adjoint  $H^\dagger$  of  $H$ given by
\begin{equation}
H^\dagger = H^\dagger_0+ \frac{1}{\sqrt{\epsilon}}H^\dagger_1 + \frac{1}{\epsilon}H^\dagger_2~,
\end{equation}
where 
\begin{eqnarray}
H^\dagger_0f&=& \frac{\partial^2 f}{\partial q^2} - (v\psi'(q)-h)\frac{\partial f}{\partial q}~,\\
H^\dagger_1f&=&\frac{\mu_{12}}{\mu_{11}}\left(-u\frac{\partial f}{\partial q} + 2\frac{\partial^2f
}{\partial q \partial u} -[v\psi'(q) -h]\frac{\partial f}{\partial u} )\right)~,\\
H^\dagger_2f&=& \frac{\mu_{22}}{\mu_{11}}\left(\frac{\partial^2 f}{\partial u^2}-u\frac{\partial f}{\partial u}\right)~.
\end{eqnarray}
We proceed by computing the average velocity, denoted here by $V$,  of $q$ due to the applied field $h$ for a tracer particle which is started with the local equilibrium distribution within a single period of the potential. This  is given by
\begin{equation}
 V=\langle \frac{dq}{ds}\rangle = \int_{-\infty}^{\infty} dq du du_0 \int_{\Lambda} dq_0 \  q \frac{dP(q,u|q_0,u_0;s)}{dt} P_{peq}(q_0,u_0)~,
\end{equation}
where $\Lambda$ the integration range over $q_0$ is over a single period of length $1$ and $P_{peq}(q,u)$
is the equilibrium distribution for the variable $u$ and the periodised variable $q^*= q \ {\rm mod}(1)$. Using the Fokker-Planck equation ({\ref{fps}) then gives
\begin{equation}
 V= \int_{-\infty}^{\infty} dq du du_0 \int_\Lambda dq_0  q [H P(q,u|q_0,u_0;s)]P_{eq}(q_0,u_0)=\int_{-\infty}^{\infty} dq du du_0 \int_\Lambda dq_0[ H^\dagger q ] P(q,u|q_0,u_0;s)P_{peq}(q_0,u_0)~,
\end{equation}
where
\begin{equation}
H^\dagger q= -v\psi'(q)+ h-\frac{1}{\sqrt{\epsilon}}\frac{\mu_{12}}{\mu_{11}}u\label{Lq}~.
\end{equation}
We notice that $H^\dagger q $ is periodic in $q$ and so we can write 
\begin{equation}
\int_{-\infty}^{\infty}  du du_0 dq \int_\Lambda dq_0 P(q,u|q_0,u_0;s)P_{peq}(q_0,u_0)=
\int_{-\infty}^{\infty}dudu_0 \int_\Lambda dq dq_0 [ H^\dagger q ]\sum_{n=-\infty}^{\infty}P(q+n,u|q_0,u_0;s)P_{peq}(q_0,u_0)~. 
\end{equation}
However $\sum_{n=-\infty}^{\infty}P(q+n,u|q_0,u_0;s)= P_p(q,u|q_0,u_0;s)$ is simply the transition density for the periodic variable $q^*$ on $\Lambda$. Clearly $P_p(q,u|q_0,u_0;s)$ satisfies the same Fokker-Planck equation as $P$ but restricted to $q\in \Lambda$ and with periodic boundary conditions.

As we have assumed that $q^*$ starts in equilibrium we find have
\begin{equation}
\int_{-\infty}^\infty du_0 \int_\Lambda dq_0 P_p(q,u|q_0,u_0;s)P_{peq}(q_0,u_0)= P_{peq}(q_0,u_0)~,
\end{equation}
and we thus obtain the Stratonovich formula
\begin{equation}
 V= \int_{-\infty}^\infty du  \int_\Lambda dq  [H^\dagger q] P_{peq}(q,u)~,
\end{equation}
where
\begin{equation}
H P_{peq}(q,u)=0~,
\end{equation}
and has the periodic boundary condition $P_{peq}(q+1,u)= P_{peq}(q,u)$.
The Stokes-Einstein formula then gives the diffusion constant for the variable $q$ as
\begin{equation}
D_q^* = \frac{\partial V}{\partial h}= \int_{-\infty}^{\infty}  du  \int_\Lambda dq\  [\frac{\partial H^\dagger}{\partial h} q] P_{peq0}(q,u)+
[H^\dagger q] \frac{\partial}{\partial h}P_{peq}(q,u)~.\label{kubo}
\end{equation}
 In Eq.~ (\ref{kubo}) everything is evaluated at $h=0$ and the integral over $q$ is on a single period $\Lambda$ of the periodic potential $\psi$.
We now denote
\begin{equation}
\frac{\partial}{\partial h}P_{peq}(q,u)= r(q,u)~,
\end{equation}
which obeys
\begin{equation}
H r(q,u) -\frac{\partial}{\partial q}P_{peq0}(q,u)-\frac{\mu_{12}}{\mu_{11}} \frac{1}{\sqrt{\epsilon}}\frac{\partial }{\partial u}P_{peq0}(q,u)=0~,\label{pder}
\end{equation}
and  must in addition obey the integral constraint, due to the  conservation of probability,
\begin{equation}
\int dq du \frac{\partial}{\partial h}P_{peq}(q,u)=\int dq du\  r(q,u)=0~.\label{inc}
\end{equation}
Now we note that Eq.~ (\ref{Lq})
gives
\begin{equation}
\frac{\partial H^\dagger}{\partial h}q= 1~,
\end{equation}
and consequently
\begin{equation}
 D_q^* =\int_{-\infty}^\infty  du \int_\Lambda dq \ P_{peq0}(q,u)+
[H^\dagger q] r(q,u)~.
\end{equation}
This simplifies to
\begin{equation}
D_q^* =   1 
- \int dq du  \left(v\psi'(q)+\frac{1}{\sqrt{\epsilon}}\frac{\mu_{12}}{\mu_{11}}u\right)r(q,u)~.
\label{Dq}
\end{equation}
The above expression can also be derived using established results on the theory of transport processes \cite{bre1993} or using a general Kubo type formalism \cite{gue2015,gue2015b}. In principal the partial differential equation Eq.~(\ref{pder}) can be numerically solved and the integral in Eq.~(\ref{Dq}) can be evaluated using the solution to give a numerically exact result for  $D^*$ for all values of $\epsilon$ and we will show how this can be done in section (\ref{kubonum}).
Here will compute the corrections due to finite $\kappa$ perturbatively for small $\epsilon$ while assuming that the variables $\mu_{ij}$ are of the same order to validate the expansion we make. We work in the adjoint representation writing
\begin{equation}
r(q,u)=P_{peq0}(q,u)w(q,u)~,
\end{equation}
which gives
\begin{equation}
H^\dagger w(q,u) +v\psi'(q)+\frac{u\mu_{12}}{\mu_{11}\sqrt{\epsilon}}=0~,\label{eqw}
\end{equation} 
along with the integral constraint
\begin{equation}
\int_{-\infty}^{\infty} du \int_{\Lambda} dq P_{peq0}(q,u)w(q,u)=0~.
\end{equation}
Setting $h=0$ in  Eq.~ (\ref{Lq}), we see that Eq.~ (\ref{eqw})  has a solution $w(q,u)= q + A$, where $A$ is a constant. The integral constraint means that we must chose
\begin{equation}
A =-\langle q\rangle_{\rm eq}= \int_{-\infty}^{\infty} du \int_{\Lambda} dq  \ qP_{peq0}(q,u)~.
\end{equation}
However we must  ensure periodicity in $q$, we thus write
\begin{equation}
w(q,u) = q -\langle q\rangle_{\rm eq}+ s(q,u)~,
\end{equation}
where $s(q,u)$  has the   boundary condition
\begin{equation}
1+ s(1,u) = s(0,u)~,
\end{equation}
and obeys the equation
\begin{equation}
H^\dagger s(q,u)=0~.
\end{equation}

Plugging this into the formula for $D_q^*$ we find
\begin{equation}
 D_q^* = 1
- \int_{-\infty}^\infty du \int_\Lambda dq  P_{peq0}(q,u)(v\psi'(q)+\frac{1}{\sqrt{\epsilon}}\frac{\mu_{12}}{\mu_{11}}u)(q-\langle q\rangle_{\rm eq}+ s(q,u))~.
\end{equation}
The shift  due to the term $\langle q\rangle_{\rm eq}$ gives zero contribution to the integral. 
The part of the integrand containing $q$ can be integrated  by parts (the second term giving zero) to give
\begin{equation}
D_q^*= \frac{1}{Z_-}\exp[- v\psi(1)] - \int dq du  P_{peq0}(q,u)(v\psi'(q)+\frac{1}{\sqrt{\epsilon}}\frac{\mu_{12}}{\mu_{11}}u)s(q,u)~,\label{dqq}
\end{equation}
and here we have explicitly taken $\Lambda = [0,1]$, and in what follows we denote 
\begin{equation}
Z_\pm = \int_0^1 dq\ \exp(\pm v \psi(q)).
\end{equation}
The first  term in the integral on the right hand side of  Eq. (\ref{dqq}) can be integrated by parts to give the computationally useful formula
\begin{equation}
 D_q^* = -\int_{-\infty}^\infty du \int_\Lambda dq  P_{peq0}(q,u)(\frac{\partial s(q,u)}{\partial q}+\frac{1}{\sqrt{\epsilon}}\frac{\mu_{12}}{\mu_{11}}us(q,u))~.
\end{equation}
Now we look for a solution of the form 
\begin{equation}
s(q,u)= s_0(q,u) +\sqrt{\epsilon}s_1(q,u) + \epsilon s_2(q,u) + \sqrt{\epsilon}^3s_3(q,u)+ O(\epsilon^2)~.
\end{equation}
This yields, expanding  to order $\epsilon$,
\begin{eqnarray}
H^\dagger_2 s_0&=&0\\\label{h0}
H^\dagger_1 s_0 + H^\dagger_2 s_1&=&0\label{h1}\\
 H^\dagger_0 s_0 + H^\dagger_1 s_1 + H^\dagger_2 s_2\label{h2} &=&0\\
 H^\dagger_0 s_1 + H^\dagger_1 s_2 + H^\dagger_2 s_3 \label{h3}&=&0\\
 H^\dagger_0s_2 + H^\dagger_1 s_3 + H^\dagger_2 s_4 \label{h4} &=&0~.
 \end{eqnarray}
 The resulting diffusion constant is then given by
 \begin{eqnarray}
 D_q^* &=&  -\frac{1}{\sqrt{\epsilon}}\int dq du  P_{peq0}(q,u)\frac{\mu_{12}}{\mu_{11}}us_0(q,u)-\int dq du  P_{peq0}(q,u)\left[\frac{\partial s_0(q,u)}{\partial q}+\frac{\mu_{12}}{\mu_{11}}us_1(q,u)\right]\nonumber \\
&-&\sqrt{\epsilon}\int dq du  P_{peq0}(q,u)\left[\frac{\partial s_1(q,u)}{\partial q}+\frac{\mu_{12}}{\mu_{11}}us_2(q,u)\right]\nonumber \\
&-&\epsilon\int dq du  P_{peq0}(q,u)\left[\frac{\partial s_2(q,u)}{\partial q}+\frac{\mu_{12}}{\mu_{11}}us_3(q,u)\right]~,
\end{eqnarray}
and for notationally simplicity we have not explicitly written the integration ranges which are $u\in (-\infty,\infty)$ and $q\in [0,1]$.
The equation (\ref{h0}), $H^\dagger_0s_0(q,u) =0$,  shows that $s_0(q,u)=s_0(q)$ is independent of $u$. Note that in terms of the parity with respect to $u$ the operators $H^\dagger_0$ and $H^\dagger_2$ are even and $H^\dagger_1$ is odd. From the structure of the equations this means that $s_n(q,u)$ is even in $u$ for $n$ even and odd in $u$ for $n$ odd. As a consequence we find that 
\begin{equation}
D^*_q = -\int dq du  P_{peq0}(q,u)\left[\frac{\partial s_0(q,u)}{\partial q}+\frac{\mu_{12}}{\mu_{11}}us_1(q,u)\right] -\epsilon\int dq du  P_{peq0}(q,u)\left[\frac{\partial s_2(q,u)}{\partial q}+\frac{\mu_{12}}{\mu_{11}}us_3(q,u)\right]~,
\end{equation}
and it is easy to see that the higher order corrections are also integer powers of $\epsilon$. We now consider Eq.~ (\ref{h1}):
$H^\dagger_1 s_0 + H^\dagger_2 s_1 =0$.
From this we see that solution for $s_1$ exists if the Fredholm alternative condition holds (the solvability condition)
\begin{equation}
\int d u P_{eq}(u) H^\dagger_1 s_0(q) = 0~,
\end{equation}
where $P_{eq}(u)= \frac{1}{\sqrt{2\pi}}\exp(-\frac{u^2}{2})$, the marginal equilibrium probability distribution of $u$,  is in the kernel of the adjoint of $H_2$  that is to say $H^\dagger_2$.  We find that 
\begin{equation}
H^\dagger_1 s_0(q) = -\frac{\mu_{12}}{\mu_{11}}u\frac{\partial s_0(q)}{\partial q}~,
\end{equation}
and this gives
\begin{equation}
s_1(q,u) =- \frac{\mu_{12}}{\mu_{22}}u\frac{\partial s_0(q)}{\partial q}~.
\end{equation}
Now consider the equation
Eq.~ (\ref{h2}), $H^\dagger_0 s_0 + H^\dagger_1 s_1 + H^\dagger_2 s_2=0$,
here we find that
\begin{equation}
H^\dagger_1 s_1 = \frac{\mu_{12}^2}{\mu_{11}\mu_{22}}([u^2- 2] \frac{\partial^2 s_0(q)}{\partial q^2}+v\psi'(q)\frac{\partial s_0(q)}{\partial q})~.
\end{equation}
Integrating Eq.~ (\ref{h2})  with respect to $P_{eq}(u)$ then gives the solvability condition for $s_2$ as  
\begin{equation}
H^\dagger_0 s_0 + \frac{\mu_{12}^2}{\mu_{11}\mu_{22}}(- \frac{\partial^2 s_0(q)}{\partial q^2}+v\psi'(q)\frac{\partial s_0(q)}{\partial q})=0~,
\end{equation}
which is simply
\begin{equation}
(1-\frac{\mu_{12}^2}{\mu_{11}\mu_{22}})H^\dagger_0s_0 = 0~.
\end{equation}
This has the solution 
\begin{equation}
s_0(q) =A \int_0^ q \exp[v\psi(p)] dp ~, 
\end{equation}
where $A$ is a constant. Note that we can choose $s_0(0)=0$ without loss of generality as $D^*_q$ is independent of any constant term in $s(q)$. Using the boundary condition   $s_0(1)=-1$ then gives
\begin{equation}
s_0(q) = -\frac{\int_0^ q \exp[v\psi(p)] dp}{Z_+}~,
\end{equation}
and so 
\begin{equation}
s'_0(q)= -\frac{\exp[v\psi(q)]}{Z_+}~.
\end{equation}
This then yields
\begin{equation}
s_1(q,u) =\frac{\mu_{12}}{\mu_{22}}u\frac{\exp[v\psi(q)]}{Z_+}~;
\end{equation}
importantly we see that $s_1(q,u)$ is periodic in $q$ and so the solution for $s$ at order $\sqrt{\epsilon}$ obeys the full boundary condition via the term $s_0(q)$ solely, all the other corrections must consequently be periodic. This  gives the $O(0)$ contribution to $D^*_q$
\begin{equation}
D^*_q{(0)}= \frac{1}{Z_+Z_-}[1-\frac{\mu^2_{12}}{\mu_{22}\mu_{11}}]~.
\end{equation}

Returning to Eq.~ ({\ref{h2})  $H^\dagger_0 s_0 + H^\dagger_1 s_1 + H^\dagger_2 s_2=0$ we find that
\begin{equation}
\frac{\mu_{12}^2}{\mu_{11}\mu_{22}}([u^2- 1] \frac{\partial^2 s_0(q)}{\partial q^2} + H^\dagger_2 s_2=0~.
\end{equation}
This satisfies the solvability condition, and has solution
\begin{equation}
s_2(q,u) = u^2\frac{\mu_{12}^2}{2\mu^2_{22}}\frac{\partial^2 s_0(q)}{\partial q^2} + s_{20}(q)~,
\end{equation}
where $s_{20}(q)$ is undetermined (but must be periodic). Now solving Eq.~ ({\ref{h3}), $H^\dagger_0 s_1 + H^\dagger_1 s_2 + H^\dagger_2 s_3$, we find, the periodic, solution 
\begin{equation}
s_3(q,u) = -u \frac{\mu_{12}}{\mu_{22}}s'_{20}(q)+ \frac{\mu_{12}u\exp[v\psi(q)]\left(\mu_{12}^2 u^2v^2\psi'(q)^2 + v\psi''(q) [6(\mu_{11}\mu_{22} -\mu_{12}^2) - u^2\mu_{12}^2]\right)}{6 Z_+\mu_{22}^3}~.
\end{equation}
To determine $s_{20}(q)$ we use the Fredholm alternative for solvability for Eq.~ (\ref{h4}), $H^\dagger_0 s_2 + H^\dagger_1 s_3 + H^\dagger_0 s_4=0$, which yields after some lengthy algebra
\begin{equation}
s_{20}(q) = \int_0^q \exp[v\psi(p)] dp [A' + \frac{ \mu_{12}^2(v^2\psi'(p)^2-v\psi''(p))}{2 \mu_{22}^2Z_+}]~,
\end{equation}
Where we have again taken $s_{20}(0)=0$,  and so the   (periodic) boundary condition at $q=1$ is $s_{20}(1)=0$ (recalling that  the inhomogeneous boundary condition is satisfied by $s_0(q)$), we thus have 
\begin{equation}
A' Z_+ + \int_0^1 \exp[v\psi(p)] dp  \frac{\mu_{12}^2(v^2\psi'(p)^2-v\psi''(p))}{2 \mu_{22}^2Z_+}]= 0~,
\end{equation}
and so
\begin{equation}
A' =- \frac{\mu_{12}^2}{ \mu_{22}^2Z^2_+}\int_0^1 \exp[v\psi(p)] v^2\psi'(p)^2 dp~.
\end{equation}
This then yields
\begin{equation}
s_{20}'(q) = \left [ - \frac{\mu_{12}^2}{ \mu_{22}^2Z^2_+}\int_0^1 \exp[v\psi(p)] v^2\psi'(p)^2 dp + \frac{ \mu_{12}^2(v^2\psi'(p)^2-v\psi''(p))}{2 \mu_{22}^2Z_+}\right]\exp[v\psi(q)]~.
\end{equation}
From the above, we find that the  correction to $D_q^*$ to $O(\epsilon)$ is
\begin{equation}
\Delta D_{q2}^* = \epsilon v^2 \frac{(\mu_{11}\mu_{22}-\mu_{12}^2)}{\mu_{22}\mu_{11}}\frac{\mu_{12}^2}{\mu_{22}^2}\frac{\int_0^1 \exp[v\psi(p)] \psi'(p)^2 dp}{Z_- Z_+^2}~.
\end{equation}
Using this to compute $D^*$ then gives the formula  derived  from the heuristic argument presented in section (\ref{heur}).
\section{The compliant limit $\epsilon \gg 1$.}
In the limit $\epsilon \gg 1$ one can carry out a similar analysis to the limit $\epsilon\ll 1$. The Kubo type formulas derived in the previous section still hold and the equation for $s(q,u)$ is now written as 
\begin{equation}
[\epsilon H_0 + \sqrt{\epsilon}H_1 + H_2] s(q,u)~,
\end{equation}
and the formula for $D_q^*$ is unchanged.

Here we find that one need to write a perturbative expansion of the form
\begin{equation}
s(q,u) = \sqrt{\epsilon}s_0(q,u) + s_1(q,u) + \frac{1}{\sqrt{\epsilon}}s_2(q,u) + O(\frac{1}{\epsilon})~.
\end{equation}
where we find that 
\begin{eqnarray}
H_0^\dagger s_0&=&0\\
H_0^\dagger s_1 + H_1^\dagger s_0 &=& 0\\
H_0^\dagger s_2 + H_1^\dagger s_1 + H_2^\dagger s_0 &=& 0~.
\end{eqnarray}
The equation $H_0 s_0=0$ then gives $s_0(q,u) = s_0(u)$ and so to leading order we find
\begin{equation}
 D_q^* = -\int_{-\infty}^\infty du \int_\Lambda dq  P_{peq0}(q,u)(\frac{\partial s_1(q,u)}{\partial q}+\frac{\mu_{12}}{\mu_{11}}s'_0(u))+O(\frac{1}{\sqrt{\epsilon}})~.\label{do1}
\end{equation}
We impose the non-periodic boundary condition on the $O(1)$  term $s_1(q,u)$, which has the boundary conditions
\begin{equation}
s_1(0,u) =0 \ ; \ 1 + s_1(1,u) = 0~.
\end{equation}
The term $s_1$ obeys the equation $H_1 s_0 + H_0 s_1=0$, where $H_1 s_0 = -\frac{\mu_{12}}{\mu_{11}}v \psi'(q) s'(u)$. We notice that  
\begin{equation}
\int_\Lambda  dq P_{eq}(q) H_1 s_0=0~,
\end{equation}
where here
\begin{equation}
P_{eq}(q)=\frac{ \exp[-v\psi(q)]}{Z_-}~,
\end{equation}
and so this equation obeys the solvability criterion.
The solution is given by
\begin{equation}
s_1(q,u) = -q\frac{\mu_{12} s_0'(u)}{\mu_{11}}+ \frac{1}{Z_+}\left[\frac{\mu_{12} s_0'(u)}{\mu_{11}}-1\right]\int_0^q dp\ \exp[ v\psi(p)]~.
\end{equation}
Using this result in Eq. (\ref{do1}) then gives to leading order
\begin{equation}
D_q^* =\frac{1}{Z_+Z_-} \left[1 - \frac{1}{\sqrt{2\pi}}\frac{\mu_{12}}{\mu_{11}}\int_{-\infty}^\infty  du \exp(-\frac{u^2}{2}) s'_0(u)\right]+O(\frac{1}{\sqrt{\epsilon}})~.\label{intd}
\end{equation}
Finally the equation for $s_0(u)$ can be found by using the solvability condition for $s_2$ 
\begin{equation}
0= \int_\Lambda  dq\ P_{eq}(q)[ H_0^\dagger s_2 + H_1^\dagger s_1 + H_2^\dagger s_0 ] = H_2^\dagger s_0(u).
+\int_\Lambda  dq\ P_{eq}(q)H_1^\dagger s_1(q,u)~.
\end{equation}
After some algebra, the solution which is integrable with respect to the equilibrium measure  $P_{eq}(u)$ for $u$, is found to be
\begin{equation}
s_0(u) = \frac{u\mu_{12}\mu_{11}}{(\mu_{11}\mu_{22}-\mu_{12}^2)Z_+Z_-+\mu_{12}^2}~.
\end{equation}
Using this in Eq. (\ref{intd}) then gives
\begin{equation}
 D_q^*=\frac{\mu_{11}\mu_{22}-\mu_{12}^2}{Z_+Z_-(\mu_{11}\mu_{22}-\mu_{12}^2)+\mu_{12}^2}+O(\frac{1}{\sqrt{\epsilon}})~.
\end{equation}
Note that when there is no periodic potential, $v=0$, one has $Z_+=Z_-=1$ and we find
\begin{equation}
D_q^*(\epsilon=\infty)= k_BT \frac{\mu_{11}\mu_{22}-\mu_{12}^2}{\mu_{22}}= \frac{k_BT}{\gamma_{11}} 
\end{equation}
which is the same result for the diffusion constant in the non compliant case $\epsilon=0$ and with $v=0$. In terms of the
unscaled variables this then gives
\begin{equation}
 D^*=k_B T \mu_{11}\frac{\mu_{11}\mu_{22}-\mu_{12}^2}{\lambda^{-2}Z_+Z_-(\mu_{11}\mu_{22}-\mu_{12}^2)+\mu_{12}^2}+O(\frac{1}{\sqrt{\epsilon}})~.
\end{equation}

\section{Numerical computation of $D^*$ from the Kubo formula \label{kubonum}}
The diffusion constant $D_q^*$ related to $D^*$ by Eq. (\ref{dqd}) can be numerically computed by solving Eq. (\ref{pder}) with periodic boundary conditions and the integral constraint Eq. (\ref{inc}). This can be carried out by using the FlexPDE Multiphysics Software For Partial Differential Equations. The numerical domain is finite and taken over $[-1/2,1/2]$ for the variable $q$ with periodic boundary conditions. The variable $u$ is unbounded but we take a finite domain with
$u\in [-L,L]$ (typically it suffices to take $L=5$),  and impose no-flux boundary conditions at the surfaces $u=\pm L$. This means that the effective potential for $u$ becomes infinite at the edges, but in practice the variable $r(q,u)$ has decayed to zero at these values of $u$. Examples are shown in Figs.~4, ~S1, and~S2, for various parameters.
\begin{figure}[h!]
    \includegraphics[width=0.5\textwidth]{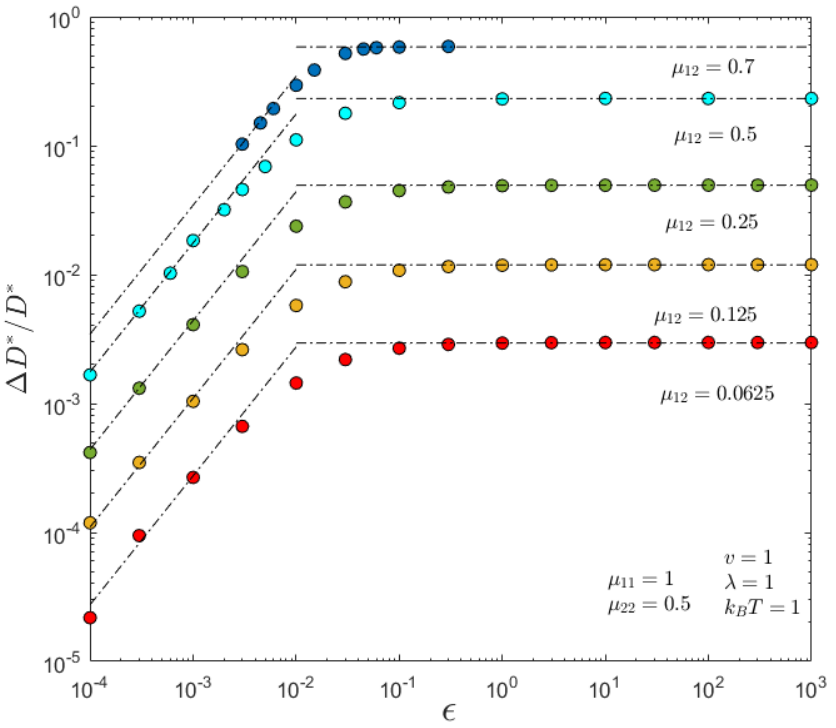}. 
   \label{fig:figS1}
   \caption{Relative increase $\Delta D^* / D^*=[D^*(\epsilon)-D^*(0)] / D^*(0)$ of the late-time diffusion coefficient $D^*$ 
    of the test particle as a function of the dimensionless compliance $\epsilon$, for \(\mu _{11}\)= 1.0, \(\mu _{22}\) = 0.5,
    \(v\) = 1.0, \(\lambda\) = 1.0, \(k_\mathrm{B}T\) = 1.0, and various \(\mu_{12}\) as indicated. The symbols correspond to the Kubo formula of Eq.~(S65). The dash-dotted 
      lines correspond to the small-$\epsilon$ prediction of Eq.~(5), and the large-$\epsilon$ prediction of Eq.~(11).}
\end{figure}
\begin{figure}[h!]
    \includegraphics[width=0.5\textwidth]{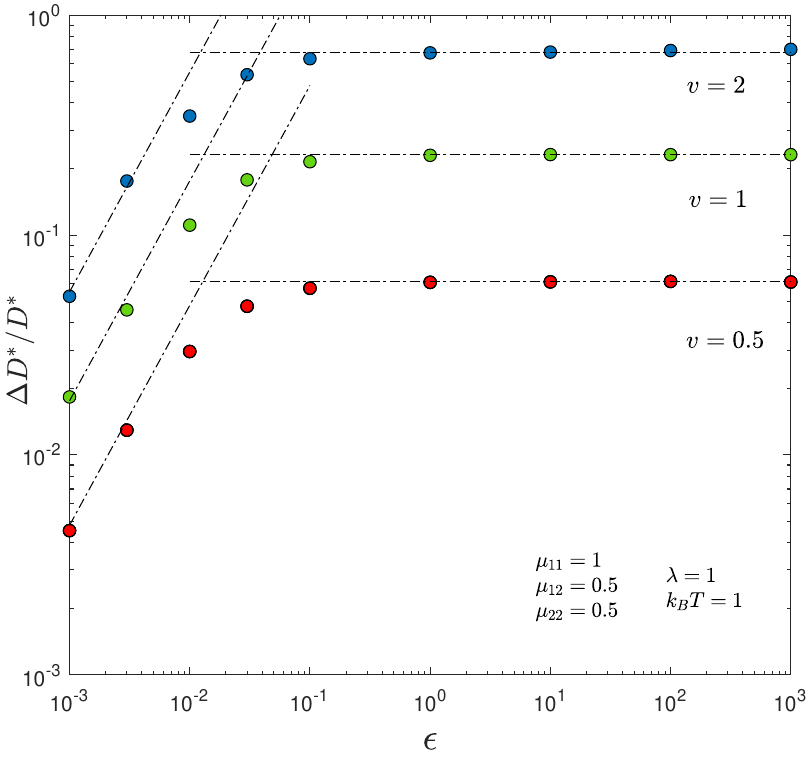}. 
   \label{fig:figS2}
    \caption{Relative increase $\Delta D^* / D^*=[D^*(\epsilon)-D^*(0)] / D^*(0)$ of the late-time diffusion coefficient $D^*$ 
    of the test particle as a function of the dimensionless compliance $\epsilon$, for \(\mu _{11}\)= 1.0, \(\mu _{12}\) = 0.5,
   \(\mu _{22}\) = 0.5, \(\lambda\) = 1.0, \(k_\mathrm{B}T\) = 1.0, and various \(v\) as indicated. The symbols correspond to the Kubo formula of Eq.~(S65). The dash-dotted 
      lines correspond to the small-$\epsilon$ prediction of Eq.~(5), and the large-$\epsilon$ prediction of Eq.~(11).}
\end{figure}

\section{Heuristic argument for modification of diffusion with inertia}
Here, we show how the heuristic method proposed beforehand can be used to compute the effective diffusion 
constant for a particle due to first order inertial effects, recovering the result of~\cite{pav2008,hai2008}. 
This calculation also demonstrates how a periodic potential can be applied to detect the signature of inertia on late time 
diffusion.

We consider the zero-temperature equation
\begin{equation}
  m\frac{\mathrm{d}v}{\mathrm{d}t} + \gamma v= -\phi'(q)~,
  \end{equation}

\noindent where $v= \frac{\mathrm{d}q}{\mathrm{d}t}$ is the particle's velocity.

In operator notation, this formally gives
\begin{equation}
v(t) =- \frac{1}{\gamma}\left(1 +\tau\frac{\mathrm{d}}{\mathrm{d}t}\right)^{-1}\phi'(q)~,
\end{equation}

\noindent where $\tau = \frac{m}{\gamma}$ is the short timescale over which the velocity relaxes and 
which would be difficult to resolve experimentally.

Now, expanding to first order in $\tau$, we find
\begin{equation}
v(t) =- \frac{1}{\gamma}\phi'(q) + \frac{\tau}{\gamma}\phi''(q) v(t)~,
\end{equation}

\noindent and so
\begin{equation}
\frac{dq}{dt}= v(t) = -\mu_\mathrm{e}(q)\phi'(q)~,
\end{equation}

\noindent where
\begin{equation}
\mu_\mathrm{e}(q) = \frac{1}{\gamma \left[1- \frac{\tau}{\gamma}\phi''(q)\right]}~.
\end{equation}

The diffusion constant is thus
\begin{equation}
D_\mathrm{e}(q) = \frac{k_\mathrm{B}T}{\gamma \left[1- \frac{\tau}{\gamma}\phi''(q)\right]}~.
\end{equation}

Now, applying the Lifson-Jackson formula, we obtain
\begin{equation}
D^*= \frac{k_\mathrm{B}T\lambda^2}{\gamma \int_0^\lambda 
\mathrm{d}q \exp[\beta \phi(q)]\left[1- \frac{m}{\gamma^2}\phi''(q)\right] \int_0^\lambda \mathrm{d}q \exp[-\beta \phi(q)]}~.
\end{equation}

This can be rewritten as
\begin{equation}
D^*= \frac{k_\mathrm{B}T\lambda^2}{\gamma 
\int_0^\lambda dq \exp[\beta \phi(q)]\left[1+\frac{m}{k_\mathrm{B}T\gamma^2}\phi'(q)^2\right]
 \int_0^\lambda \mathrm{d}q \exp[-\beta \phi(q)]}~,
\end{equation}

\noindent or to the same order
\begin{equation}
D^*= \frac{k_\mathrm{B}T\lambda^2}{\gamma \int_0^\lambda \mathrm{d}q \exp[\beta \phi(q)]
 \int_0^\lambda \mathrm{d}q \exp[-\beta \phi(q)]}\left(1
-\frac{m}{k_\mathrm{B}T\gamma^2}\frac{\int_0^\lambda \mathrm{d}q \exp[\beta \phi(q)]
\phi'(q)^2}{\int_0^\lambda \mathrm{d}q \exp[\beta \phi(q)]}\right)~,
\end{equation}
\noindent which recovers the result given in~\cite{pav2008,hai2008} via multiscale analysis.
 Notice that in this particular case, the effect of inertia is to reduce the late time diffusion constant,
  and again, it is necessary to impose an external periodic potential to produce a measurable effect on the late time diffusivity.

  \section{Numerical Simulations}

  To study the dynamics of the system, we numerically solve the discretized Langevin equations using the mobility tensor formulation. 
  Employing the Euler-Maruyama scheme \cite{maruyama_continuous_1955}, the equations are given by
  
  \begin{eqnarray} 
    q_1(t+\Delta t) - q_1(t) &=& -\mu_{11} \phi'(q_1)\Delta t - \frac{\mu_{12}}{\kappa} q_2 \Delta t + \xi_1 \sqrt{2 k_\mathrm{B}T \Delta t}, \label{dq1}\\
    q_2(t+\Delta t) - q_2(t) &=& -\frac{\mu_{22}}{\kappa} q_2 \Delta t - \mu_{12} \phi'(q_1)\Delta t + \xi_2 \sqrt{2 k_\mathrm{B}T \Delta t},\label{dq2} 
  \end{eqnarray}

  \noindent where $\xi_{i}$ represent zero-mean Gaussian random variables, characterised by the correlation
  \begin{equation} 
    \langle \xi_{i} \xi_{j} \rangle = \mu_{ij}~. 
  \end{equation}
  
  The mobility tensor $\mu_{ij}$ is required to be a positive-definite matrix. To enforce this, we write $\xi_i$ 
  in terms of the variables $\eta_{j}$ as
  
  \begin{equation} 
    \xi_{i} = m_{ij}\eta_{j}~, \label{xim} 
  \end{equation}
  
  \noindent where $\eta_{i}$ are zero-mean, independent Gaussian random variables satisfying 
  $\langle \eta_\mathrm{i} \eta_{j} \rangle = \delta_{ij}$. This yields a decomposition of the mobility tensor
  
  \begin{equation} 
    \mu_{ij} = m_{ik} m_{kj}~. 
  \end{equation}
  
  The matrix $m_\mathrm{ij}$ is constructed to produce the desired mobility tensor $\mu_\mathrm{ij}$, 
  ensuring that noise terms are appropriately correlated. For computational efficiency, the noise array is pre-generated 
  using the \texttt{numpy.random.normal} function within the framework of Eq.~(\ref{xim}). The simulation
  recursively updates the system state at each time step.

  We carefully choose the time step $\Delta t$ for numerical stability and accuracy. A step size that is too large may yield erroneous 
  convergence of statistical quantities, while an excessively small step size can hinder computational efficiency and reduce the 
  ability to gather sufficient statistics at long times. To ensure robustness, we restrict $\Delta t$ to values satisfying 
  $\Delta t \leq 10^{-2} \tau_2$, where the characteristic time $\tau_2$ is defined as $\tau_2 = \kappa \gamma_{22}$.
   This choice maintains the system within the overdamped regime and ensures accurate short-time statistics.
  
  Trajectories are computed iteratively, and their corresponding PDFs for displacements are evaluated. The PDFs are stacked, and the root mean square error 
  (RMSE) between simulation results and the theoretical predictions from Eq.~(\ref{gbd}) is computed. The simulations are iterated until the RMSE 
  drops below a threshold of 1\%, ensuring statistical consistency.

  The modified long-time diffusion coefficient is extracted from the mean squared displacement (MSD) of $q_1$ by performing a log-log fit of the form:
  
\begin{equation}
  \mathrm{ln}(\mathrm{MSD})=\mathrm{ln}(\mathrm{A})+\mathrm{B}\mathrm{ln}(t)~.
\end{equation}
  
  To ensure uniform statistical weighting, this fit is conducted over points in log space, minimising biases toward higher-time values. 
  We exclude data points corresponding to time scales before $\tau^* = \lambda^2 / D^*$, which belong to a different dynamical regime where linear diffusion has not yet set in. The time scale $\tau^*$ corresponds to the time that the tracer spends in a single period of the potential, trapped in a local minimum. 
 We also neglect  points in the last decade of the simulation, where the statistics are less reliable due to less data points being available.
  
  For additional robustness, multiple fits are performed using averaged MSDs computed from different simulated trajectories of a same subset of parameters.
   The diffusion coefficient is only accepted when the coefficients of the linear regression become consistent across new iterations to confirm the convergence.
  
  The simulation framework scripts are fully open-source and available on our GitHub repository \url{https://github.com/EMetBrown-Lab/Lacherez2024\textunderscore Codes}. 
  The repository provides a \emph{Jupyter} 
  notebook for single-trajectory simulations to explore the system with minimal setup. 
  For large-scale parameter sweeps, the \texttt{setup.py} script enables optimised pre-compilation using \emph{Cython}, and the 
  \texttt{automate\_simulation.py} script supports systematic exploration of parameter spaces. These tools are designed for efficient single-node
   execution and ensure reproducibility of all results presented in this study. Additional treatment code is available upon reasonable request to the authors.